\documentclass[amsmath,amssymb,aps,floatfix,prd,a4paper,superscriptaddress]{revtex4}

\usepackage{bm}
\usepackage{amsfonts}
\usepackage{booktabs}
\usepackage{float}
\usepackage{graphicx}
\usepackage{tikz}
\usepackage[colorlinks=true,linkcolor=blue,citecolor=blue,urlcolor=blue]{hyperref}

\tikzset{
  hadron/.style={draw=black,line width=0.55pt},
  fdot/.style={circle,fill=black,inner sep=1.15pt}
}

\newcommand{\treemechanism}{%
\begin{tikzpicture}[x=0.84cm,y=0.62cm,baseline=(current bounding box.center),
  every node/.style={font=\scriptsize}]
  \node[fdot] (v) at (0,0) {};
  \draw[hadron] (-1.25,0)--(v) node[pos=0.32,above] {$\psi$};
  \draw[hadron] (v)--(1.55,1.15) node[right] {$\bar\Xi^0$};
  \draw[hadron] (v)--(1.75,0.38) node[right] {$p$};
  \draw[hadron] (v)--(1.75,-0.38) node[right] {$K^-$};
  \draw[hadron] (v)--(1.55,-1.15) node[right] {$\bar K^0$};
\end{tikzpicture}}

\newcommand{\fsimechanism}[5]{%
\begin{tikzpicture}[x=0.82cm,y=0.60cm,baseline=(current bounding box.center),
  every node/.style={font=\scriptsize}]
  \node[fdot] (a) at (0,0) {};
  \node[fdot] (b) at (1.65,0) {};
  \draw[hadron] (-1.20,0)--(a) node[pos=0.30,above] {$\psi$};
  \draw[hadron] (a)--(0.52,1.18) node[above] {$\bar\Xi^0$};
  \draw[hadron] (a)--(0.52,-1.18) node[below] {$#3$};
  \draw[hadron] (a) .. controls (0.45,0.56) and (1.20,0.56) ..
    (b) node[pos=0.50,above] {$#2$};
  \draw[hadron] (a) .. controls (0.45,-0.56) and (1.20,-0.56) ..
    (b) node[pos=0.50,below] {$#1$};
  \draw[hadron] (b)--(2.68,0.55) node[right] {$#5$};
  \draw[hadron] (b)--(2.68,-0.55) node[right] {$#4$};
\end{tikzpicture}}

\newcommand{\be}{\begin{equation}}
\newcommand{\ee}{\end{equation}}
\newcommand{\beqa}{\begin{eqnarray}}
\newcommand{\eeqa}{\end{eqnarray}}

\begin{document}

\title{Meson-baryon final-state interactions in the
$\psi(3686)\to pK^-K_S^0\,\bar\Xi^0$ decay}

\author{Fernando C\'esar Sobrinho}
\affiliation{Instituto de F\'\i{}sica, Universidade de S\~ao Paulo,
Rua do Mat\~ao 1371, CEP 05508-090, S\~ao Paulo, SP, Brazil}
\affiliation{Departamento de F\'\i{}sica Te\'orica and IFIC,
Centro Mixto Universidad de Valencia-CSIC,
Institutos de Investigaci\'on de Paterna,
46071 Valencia, Spain}

\author{Wen-Tao Lyu}
\affiliation{School of Physics, Zhengzhou University,
Zhengzhou 450001, Henan, China}
\affiliation{Departamento de F\'\i{}sica Te\'orica and IFIC,
Centro Mixto Universidad de Valencia-CSIC,
Institutos de Investigaci\'on de Paterna,
46071 Valencia, Spain}

\author{Eulogio Oset}
\affiliation{Departamento de F\'\i{}sica Te\'orica and IFIC,
Centro Mixto Universidad de Valencia-CSIC,
Institutos de Investigaci\'on de Paterna,
46071 Valencia, Spain}

\begin{abstract}
We study the decay $\psi(3686)\to p K^- \bar K^0 \bar\Xi^0$, treating the $\psi(3686)$ as a $c\bar c$ state and therefore as a singlet under SU(3) transformations in the light-quark sector. We consider all possible combinations of a baryon, an antibaryon, and two pseudoscalar mesons that form an SU(3) singlet, retaining the terms that can lead to the $p K^- \bar K^0 \bar\Xi^0$ final state through $S$-wave final-state interactions. Three independent SU(3)-singlet structures are identified and assigned different weights. We also include the contribution of the $\Lambda(1520)$ in the meson–baryon $D$-wave. A fit to the data provides a good description of all six invariant-mass distributions. In addition to the $\Lambda(1520)$ contribution, we find a small contribution from the $\Lambda(1670)$ in the $K^-p$ invariant-mass distribution, generated by final-state interactions, and a contribution from the $\Sigma(1430)$, which produces a significant distortion of the $\bar K^0p$ invariant-mass distribution relative to phase space. This study demonstrates the value of investigating charmonium decays into a baryon–antibaryon pair and two pseudoscalar mesons, complementing the more commonly studied decays into a baryon–antibaryon pair and a single pseudoscalar meson.
\end{abstract}

\maketitle

\section{Introduction}\label{sec1}

Hyperon resonances provide an important testing ground for understanding hadron structure and the nonperturbative dynamics of quantum chromodynamics (QCD) at low energies~\cite{Wang:2024jyk}. Although the ground-state baryon octet and decuplet are well established, the existence and nature of several low-lying excited baryons remain under debate, particularly in the sector with $I(J^P)=1(1/2^-)$. The development of baryonic chiral Lagrangians~\cite{Ecker:1994gg} and chiral unitary approaches~\cite{Oset:1997it,Oller:2000fj,Lutz:2001yb,Jido:2003cb,Hyodo:2011ur,Jido:2001nt,Garcia-Recio:2002yxy} has provided new insights into these states. In this framework, some baryon resonances emerge dynamically from coupled-channel meson--baryon interactions, offering an interpretation beyond the conventional three-quark picture. A prominent example is the two-pole structure associated with the $\Lambda(1405)$, with poles typically located near 1380 and 1420~MeV, depending on the theoretical analysis~\cite{Oller:2000fj,Jido:2003cb}. The $\Lambda(1670)$ has also been interpreted as a dynamically generated resonance arising from the coupled-channel interactions of $\bar{K}N$, $\pi\Sigma$, $\eta\Lambda$, and $K\Xi$ in the isospin-zero sector~\cite{Oset:1997it,Oset:2001cn}.

The isospin-one sector is less well established. The PDG lists the $\Sigma(1620)$ as a candidate with $J^P=1/2^-$, but its experimental status remains uncertain~\cite{ParticleDataGroup:2026mpi}. Meanwhile, chiral unitary analyses of $S$-wave pseudoscalar-meson--octet-baryon scattering in the strangeness $S=-1$ sector predict a lower-mass state with $I(J^P)=1(1/2^-)$, commonly referred to as the $\Sigma(1430)$, near the $\bar{K}N$ threshold~\cite{Oset:2001cn,Oset:1997it,Khemchandani:2018amu,Kamiya:2016jqc,Jido:2003cb,Oller:2006jw,Garcia-Recio:2002yxy,Lutz:2001yb}. In 2022, the Belle Collaboration studied the $\Lambda_c^+\to\Lambda\pi^+\pi^+\pi^-$ reaction and reported evidence for a structure in the $\pi^+\Lambda$ and $\pi^-\Lambda$ mass distributions near the $\bar{K}N$ threshold~\cite{Belle:2022ywa}, where the $\Sigma$ state had been predicted, further motivating investigations of its nature.

Charmonium decays offer a useful environment for investigating final-state interactions and the dynamical origin of baryon resonances. Considerable theoretical effort has been devoted to exploring exotic hadronic states in these processes~\cite{Wang:2020wap,Ding:2023eps,Ding:2024lqk,Ren:2026ila,Lyu:2026ums,Dai:2026zqn,Lyu:2026rsm,He:2026mkf,Ikeno:2026zpv,Song:2026gnn}. Beyond three-body decays, four-body final states are frequently observed in scattering reactions and particle decays~\cite{BESIII:2025eqp,BESIII:2024ziy,LHCb:2022orj,BESIII:2017jyh,LHCb:2016qbq,ANKE:2007dyc}. Their theoretical description is more involved because of the richer kinematics and the possible interplay among different final-state interactions. Consequently, these processes have received comparatively limited theoretical attention. Some theoretical studies include the reactions $pp\to pp\pi^+\pi^-$, $pp\to pK^+K^-p$, and $pp\to pK^+\pi^0\Sigma^0$ in Refs.~\cite{Lebiedowicz:2009pj,Xie:2010md,Lu:2014rla}, respectively, as well as the decays $D_s^+\to\pi^+\pi^+\pi^-\eta$ in Ref.~\cite{Song:2022kac}, $\Lambda_c^+\to\Lambda\pi^+\pi^+\pi^-$ in Refs.~\cite{He:2025vij,Li:2025yad}, $D_s^+\to\pi^+\pi^0\pi^0\eta$ in Ref.~\cite{He:2026iiy}, and $D^+\to\bar{K}^0\pi^+\pi^0\pi^0$ in Ref.~\cite{Yang:2026bhm}.

Recently, the BESIII Collaboration observed the decay $\psi(3686)\to pK^-K_S^0\bar{\Xi}^0+\mathrm{c.c.}$ and measured its branching fraction using a sample of $(2.712\pm0.014)\times10^9$ $\psi(3686)$ events~\cite{BESIII:2026qzk}. A clear structure near 1520~MeV was observed in the $K^-p$ invariant mass distribution and attributed to the $\Lambda(1520)$. The theoretical analysis of this four-body decay is challenging because of its multidimensional phase space and the possible interplay among different final-state interactions. At the same time, the six two-body subsystems provide more opportunities to investigate final-state interactions than those available in a three-body decay. Their invariant mass distributions offer complementary information on possible resonant structures, making this process a promising setting for studying the dynamics of hyperon resonances.

In this work, we investigate the decay $\psi(3686)\to pK^-\bar{K}^0\bar{\Xi}^0$ by incorporating the $K^-p$ and $\bar{K}^0p$ final-state interactions within a chiral unitary framework. The coupled-channel meson--baryon interactions dynamically generate the two-pole structure associated with the $\Lambda(1405)$ in the isospin-zero sector. They also generate the $\Lambda(1670)$ in the same sector and the $\Sigma(1430)$ in the isospin-one sector. We examine their effects on the line shapes of the two-body invariant mass distributions and also include the contribution of the $\Lambda(1520)$. We fit the available experimental data and calculate all six two-body invariant mass distributions, assessing how well the model describes the measured spectra.

The remainder of this paper is organized as follows. In Sec.~\ref{sec2}, we present the theoretical formalism for the reactions $\psi(3686)\to pK^-\bar{K}^0\bar{\Xi}^0$. The numerical results and corresponding discussions are given in Sec.~\ref{sec3}. Finally, a brief summary is provided in Sec.~\ref{sec4}.

\section{Formalism}
\label{sec:formalism}
\label{sec2}

\subsection{Physical process and approximations}
\label{subsec:scope}

The strong process considered in this work is
\be
\psi(3686)(P_\psi,\epsilon)
\longrightarrow
p(p_p)\,K^-(p_{K^-})\,\bar K^0(p_{\bar K^0})\,
\bar\Xi^0(p_{\bar\Xi^0}),
\label{eq:physical_process}
\ee
where the produced $\bar K^0$ is observed experimentally through its
$K_S^0$ component~\cite{BESIII:2026qzk}. The first consideration is that, in order to have this final state, it is not
necessary to produce it directly. The decay can proceed producing other
states that, through final-state interaction, can lead to the desired final
state. In this respect, we will consider the coupled channels that through
interaction can produce resonances and hence contribute effectively to the
decay of Eq.~(\ref{eq:physical_process}).

In this sense, the $K^-\bar K^0$ pair has strangeness $S=-2$, and there are no
meson resonances with $S=-2$. Actually, lattice-QCD calculations give a very
small scattering length for the $K^+K^+$ interaction,
$
a\simeq 0.14\ \mathrm{fm},
$
the charge conjugate of $K^-K^-$. For this reason, we shall not consider the
interaction of the $K^-\bar K^0$ pair~\cite{Beane:2008}.

On the other hand, the $K^-\bar\Xi^0$ or $\bar K^0\bar\Xi^0$ interaction,
equivalent to that of $K\Xi$, corresponds to a coupled channel to $\bar K N$
and other channels that generate the two $\Lambda(1405)$ states, the
$\Lambda(1670)$, and two $\Sigma$ states, $\Sigma(1430)$ and
$\Sigma(1620)$~\cite{Li:2025yad,Ren:2026ila}.
However, the threshold of $K\Xi$, at about $1808\ \mathrm{MeV}$, is far away
from the highest-mass generated resonance, the $\Lambda(1670)$. Thus, the
consideration of the $K\Xi$ interaction, necessarily at higher energy than its
threshold, will not have any relevant repercussion in the final-state
interaction. For this reason, we also ignore this interaction, and the
$\bar\Xi^0$ will act as a spectator in our approach.

On the other hand, the strength of the interaction between baryons is weaker
than that of the meson--baryon interaction, and we also neglect the
baryon--antibaryon final-state interaction here. This has also been assumed
in former related works on charmonium going to a baryon--antibaryon pair and
pseudoscalar mesons, where the baryon--antibaryon mass distribution is very
well reproduced considering only the meson--baryon interaction. Thus, we
have only to consider the interaction of the pseudoscalar mesons with the
baryons~\cite{Dai:2026zqn,Song:2026gnn}.

We separate the reaction mechanism into the primary production, the
flavor weights, the spin structure and the final-state interaction. Then the
decay amplitude can be represented schematically as
\be
{\cal M}_f(M)=
{\cal S}_{\rm LO}
\left[h_f+\sum_i h_iG_i(M)T_{if}(M)\right].
\label{eq:schematic_amplitude}
\ee
where ${\cal S}_{\rm LO}$ stands for the spin structure of the amplitude, the
$h_i$ represent weights for the different mechanisms, $G_i$ are meson-baryon loop
functions, and $T_{if}$ are scattering matrices. All these will be described
later.

\begin{figure}[t]
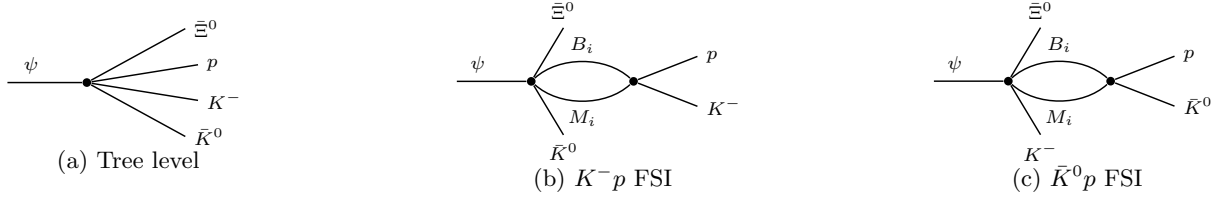

\centering
\begin{minipage}{0.30\linewidth}
\centering
\treemechanism\\[-2pt]
{\small (a) Tree level}
\end{minipage}\hfill
\begin{minipage}{0.30\linewidth}
\centering
\fsimechanism{M_i}{B_i}{\bar K^0}{K^-}{p}\\[-2pt]
{\small (b) $K^-p$ FSI}
\end{minipage}\hfill
\begin{minipage}{0.30\linewidth}
\centering
\fsimechanism{M_i}{B_i}{K^-}{\bar K^0}{p}\\[-2pt]
{\small (c) $\bar K^0p$ FSI}
\end{minipage}
\caption{Diagrammatic mechanisms included in the amplitude. The black dots
denote primary production or rescattering vertices.}
\label{fig:mechanisms}
\end{figure}

The three $S$-wave topologies retained in the diagrammatic representation are displayed in
Fig.~\ref{fig:mechanisms}. The first one is the contact tree-level production.
The following two represent the final-state meson-baryon interaction leading to either antikaon
with the proton. The loop topologies follow the diagrammatic organization used
in Refs.~\cite{Duan:2024,Ikeno:2026zpv,Lyu:2026rsm}.

The construction below assumes that the strong production conserves parity
and the primary vertex is truncated at leading order in the final
three-momenta in the $\psi(3686)$ rest frame. The
meson--baryon interaction is restricted to $S$-wave and it is unitarized with
on-shell factorization. This sector can generate negative-parity spin-$1/2$
hyperons, but it cannot generate the $J^P=3/2^-$ $\Lambda(1520)$. For this
reason the latter resonance will be introduced through a separate amplitude.

\subsection{SU(3) structure of the primary decay}
\label{subsec:flavor}

We begin with the flavor structure of the process
$\psi(3686)\to\bar B BPP$, where $B$, $\bar B$ and $P$ denote the SU(3)
matrices of baryons, antibaryons and pseudoscalar mesons, respectively. We use
the convention of Refs.~\cite{Ikeno:2026zpv,Lyu:2026rsm}, with the physical $\eta$
component of $P$ written in the $\eta$--$\eta'$ mixing convention of
Ref.~\cite{Bramon:1992},
\be
P=\left(\begin{array}{ccc}
\frac{\pi^0}{\sqrt2}+\frac{\eta}{\sqrt3}&\pi^+&K^+\\
\pi^-&-\frac{\pi^0}{\sqrt2}+\frac{\eta}{\sqrt3}&K^0\\
K^-&\bar K^0&-\frac{\eta}{\sqrt3}
\end{array}\right),
\label{eq:Pmatrix}
\ee
\be
B=\left(\begin{array}{ccc}
\frac{\Sigma^0}{\sqrt2}+\frac{\Lambda}{\sqrt6}&\Sigma^+&p\\
\Sigma^-&-\frac{\Sigma^0}{\sqrt2}+\frac{\Lambda}{\sqrt6}&n\\
\Xi^-&\Xi^0&-\frac{2\Lambda}{\sqrt6}
\end{array}\right),
\label{eq:Bmatrix}
\ee
and
\be
\bar B=\left(\begin{array}{ccc}
\frac{\bar\Sigma^0}{\sqrt2}+\frac{\bar\Lambda}{\sqrt6}
&\bar\Sigma^+&\bar\Xi^+\\
\bar\Sigma^-&-\frac{\bar\Sigma^0}{\sqrt2}
+\frac{\bar\Lambda}{\sqrt6}&\bar\Xi^0\\
\bar p&\bar n&-\frac{2\bar\Lambda}{\sqrt6}
\end{array}\right).
\label{eq:Bbar_matrix}
\ee

The $\psi(3686)$ is a $c\bar c$ state and is therefore a singlet under
light-flavor SU(3). The primary vertex must consequently be an SU(3)
scalar. We therefore work at leading single-trace order, as in
Refs.~\cite{Ikeno:2026zpv,Lyu:2026rsm}, and neglect multi-trace structures, which are suppressed by large-$N_c$ counting~\cite{Manohar:1998}, and because the traces of $B$ and $\bar B$ are null, and so is that of $P$ if the $\eta$ is considered as the octet of SU(3) ignoring the $\eta$--$\eta'$ mixing implicit in Eq.~(\ref{eq:Pmatrix}). With one baryon, one antibaryon and two pseudoscalar matrices, the three
single-trace orderings are
\be
F_1=\langle\bar BPPB\rangle,\qquad
F_2=\langle\bar BPBP\rangle,\qquad
F_3=\langle\bar BBPP\rangle.
\label{eq:traces}
\ee
In explicit matrix notation these expressions are
\beqa
F_1&=&\sum_{i,j,k,l}\bar B_{ij}P_{jk}P_{kl}B_{li},
\nonumber\\
F_2&=&\sum_{i,j,k,l}\bar B_{ij}P_{jk}B_{kl}P_{li},
\nonumber\\
F_3&=&\sum_{i,j,k,l}\bar B_{ij}B_{jk}P_{kl}P_{li}.
\label{eq:trace_indices}
\eeqa
We assign an independent strength to each structure,
\be
{\cal C}_{\rm flavor}=\widetilde A F_1+
\widetilde B F_2+
\widetilde C F_3.
\label{eq:flavor_total}
\ee
One common factor can always be absorbed into the normalization of the decay,
such that two ratios determine the relative flavor composition.

We shall use Eq.~(\ref{eq:flavor_total}) as a source of the tree level
and of the interaction of meson--baryon pairs leading to $K^-p$ in the final
state. Fixing $\bar K^0\bar\Xi^0$ as the spectator pair, this source is
\begin{align}
H_{\bar K^0\bar\Xi^0}={}&
\widetilde B K^-p+\widetilde B\bar K^0n
+\widetilde B\pi^+\Sigma^-+\widetilde A\pi^-\Sigma^+
+\frac{\widetilde A+\widetilde B}{2}\pi^0\Sigma^0
\nonumber\\
&+\frac{\widetilde B}{\sqrt2}\eta\Lambda
+\frac{2\widetilde C-\widetilde A-\widetilde B}{2\sqrt3}
\pi^0\Lambda
-\frac{\widetilde B}{\sqrt6}\eta\Sigma^0
\nonumber\\
&+(\widetilde A+\widetilde C)K^0\Xi^0
+\widetilde C K^+\Xi^-.
\label{eq:physical_polynomial}
\end{align}
The first term in Eq.~(\ref{eq:physical_polynomial}), accounting for the tree
level, is
\be
\widetilde B\,\bar\Xi^0\bar K^0K^-p.
\ee
This term can also lead to rescattering through
\be
K^-p\longrightarrow K^-p.
\ee
The other terms in Eq.~(\ref{eq:physical_polynomial}) also lead to $K^-p$
through processes such as
\be
\bar K^0n\longrightarrow K^-p,
\ee
and the remaining coupled channels indicated there. For a final $K^-p$ pair the nonvanishing production weights of Eq.~(\ref{eq:schematic_amplitude}) are therefore
\beqa
h_{K^-p}^{(-)}&=&\widetilde B,
\nonumber\\
h_{\bar K^0n}^{(-)}&=&2\widetilde B,
\nonumber\\
h_{\pi^+\Sigma^-}^{(-)}=\widetilde B,\qquad
h_{\pi^-\Sigma^+}^{(-)}&=&\widetilde A,
\nonumber\\
h_{\pi^0\Sigma^0}^{(-)}&=&\frac{\widetilde A+\widetilde B}{2},
\nonumber\\
h_{\pi^0\Lambda}^{(-)}&=&
\frac{2\widetilde C-\widetilde A-\widetilde B}{2\sqrt3},
\nonumber\\
h_{\eta\Sigma^0}^{(-)}=-\frac{\widetilde B}{\sqrt6},\qquad
h_{\eta\Lambda}^{(-)}&=&\frac{\widetilde B}{\sqrt2}.
\label{eq:weights_neutral}
\eeqa
The coefficient $h_{\bar K^0n}^{(-)}$ has a factor 2 with respect to
Eq.~(\ref{eq:physical_polynomial}) because we have two identical $\bar K^0$
mesons.

Similarly to Eq.~(\ref{eq:physical_polynomial}), we can obtain from the traces
in Eq.~(\ref{eq:traces}) the terms that isolate the $\bar\Xi^0K^-$ spectator
pair. Then we find
\begin{align}
H_{K^-\bar\Xi^0}={}&
\widetilde B\,\bar K^0p
+\frac{\widetilde B-\widetilde A}{\sqrt2}\,\pi^+\Sigma^0
+\frac{\widetilde A-\widetilde B}{\sqrt2}\,\pi^0\Sigma^+
\nonumber\\
&+\frac{\widetilde A+\widetilde B-2\widetilde C}{\sqrt6}\,
\pi^+\Lambda
+\frac{\widetilde B}{\sqrt3}\,\eta\Sigma^+.
\label{eq:charged_physical_polynomial}
\end{align}
Note that the first term in Eq.~(\ref{eq:charged_physical_polynomial}) is the
same as the first term in Eq.~(\ref{eq:physical_polynomial}). It goes one time
into the tree level, but it goes through $K^-p\to K^-p$ or
$\bar K^0p\to\bar K^0p$ in the final-state interaction of the two topologies
described in Fig.~\ref{fig:mechanisms}.

We can now obtain the weights entering the $h_iG_iT_{i,\bar K^0p}$ term of
Eq.~(\ref{eq:schematic_amplitude}). They are
\beqa
h_{\bar K^0p}^{(0)}&=&\widetilde B,
\nonumber\\
h_{\pi^+\Sigma^0}^{(0)}&=&\frac{\widetilde B-\widetilde A}{\sqrt2},
\qquad
h_{\pi^0\Sigma^+}^{(0)}=\frac{\widetilde A-\widetilde B}{\sqrt2},
\nonumber\\
h_{\pi^+\Lambda}^{(0)}&=&
\frac{\widetilde A+\widetilde B-2\widetilde C}{\sqrt6},
\qquad
h_{\eta\Sigma^+}^{(0)}=\frac{\widetilde B}{\sqrt3}.
\label{eq:weights_charged}
\eeqa

\subsection{Lorentz and spin structure}
\label{subsec:spin}

The production amplitude can be written in a general form as
\be
{\cal M}=\epsilon_\mu(P_\psi)\,
\bar u(p_p)\gamma^\mu v(p_{\bar\Xi^0}).
\label{eq:general_lorentz_amplitude}
\ee
The product of two pseudoscalar fields has positive intrinsic parity. Thus, in
contrast with the one-pseudoscalar vertex of Ref.~\cite{Ikeno:2026zpv}, that
required the $\epsilon_\mu\gamma^\mu\gamma_5$ vertex, we now need
$\epsilon_\mu\gamma^\mu$. In addition, we work in the parent rest frame, for which
$\epsilon^0=0$.

In the standard Dirac representation we have
\be
\gamma^0=\left(\begin{array}{cc}1&0\\0&-1\end{array}\right),
\qquad
\gamma^i=\left(\begin{array}{cc}0&\sigma^i\\-\sigma^i&0\end{array}\right).
\label{eq:dirac_matrices}
\ee

Keeping the first nonvanishing terms of the on-shell spinors,
\be
u(p_p)=\left(\begin{array}{c}
\chi_p\\[1mm]
\frac{\bm\sigma\cdot\bm p_p}{2m_p}\chi_p
\end{array}\right)+O(|\bm p_p|^2),
\qquad
v(p_{\bar\Xi^0})=\left(\begin{array}{c}
\frac{\bm\sigma\cdot\bm p_{\bar\Xi^0}}{2m_{\bar\Xi}}\chi_{\bar\Xi}\\[1mm]
\chi_{\bar\Xi}
\end{array}\right)+O(|\bm p_{\bar\Xi^0}|^2),
\label{eq:spinors_LO}
\ee
we find for the vector current
\be
\epsilon_\mu\bar u(p_p)\gamma^\mu v(p_{\bar\Xi^0})
=-\chi_p^\dagger
\bm\sigma\cdot\bm\epsilon\,
\chi_{\bar\Xi}+O(\bm p_p\cdot\bm p_{\bar\Xi^0}, |\bm p_p|^2, |\bm p_{\bar\Xi^0}|^2).
\label{eq:vector_reduction}
\ee
The common minus sign can be absorbed into the definition of the production
strengths. We consequently define
\be
{\cal S}_{\rm LO}=
\chi_p^\dagger\bm\sigma\cdot\bm\epsilon\,\chi_{\bar\Xi},
\qquad
{\cal M}_i^{\rm prod}=h_i{\cal S}_{\rm LO}.
\label{eq:spin_structure}
\ee
Summing the final
spins and averaging the three polarizations of the parent gives
\be
\frac{1}{3}\sum_{\lambda,s_p,s_{\bar\Xi}}
\left|\chi_p^\dagger\bm\sigma\cdot\bm\epsilon_\lambda
\chi_{\bar\Xi}\right|^2=2.
\label{eq:spin_average}
\ee

\subsection{Meson--baryon final-state interaction}
\label{subsec:fsi}

The lowest-order chiral interaction used as the source of the
coupled-channel potential is the Weinberg--Tomozawa
term~\cite{Oset:1997it,Ikeno:2026zpv}. Its low-energy $S$-wave expression is
\be
V_{ij}(W)=-\frac{C_{ij}}{4f_if_j}
\left(k_i^0+k_j^0\right),
\label{eq:potential}
\ee
where $W$ is the meson--baryon invariant mass, $C_{ij}$ are the SU(3)
coefficients, $f_i$ are the pseudoscalar decay constants and
\be
k_i^0(W)=\frac{W^2+m_i^2-M_i^2}{2W}
\label{eq:meson_energy}
\ee
is the energy of the meson in channel $i$, with $m_i$ and $M_i$ being the masses of the meson and baryon in that channel, respectively. With on-shell factorization, the
coupled-channel equation $T=V+VGT$ becomes
\be
T(W)=[1-V(W)G(W)]^{-1}V(W).
\label{eq:bethe_salpeter}
\ee
Here
\be
G(W)=\operatorname{diag}\bigl(G_i(W)\bigr).
\label{eq:diagonal_G}
\ee

The neutral sector is evaluated in the physical charge basis
\begin{align}
{\cal B}_{K^-p}=(&K^-p,\bar K^0n,\pi^0\Lambda,\pi^0\Sigma^0,
\pi^+\Sigma^-,\pi^-\Sigma^+,\nonumber\\
&\eta\Lambda,\eta\Sigma^0,K^+\Xi^-,K^0\Xi^0),
\label{eq:neutral_basis}
\end{align}
and the charged sector uses
\be
{\cal B}_{\bar K^0p}=(\bar K^0p,\pi^0\Sigma^+,\pi^+\Sigma^0,
\pi^+\Lambda,\eta\Sigma^+,K^+\Xi^0).
\label{eq:charged_basis}
\ee
The matrices $C_{ij}$ for the neutral charge basis follow
Ref.~\cite{Oset:1997it}, while the charged-sector coefficients and
phase convention follow Ref.~\cite{Li:2025yad}.

The neutral $K^-p$ sector uses the effective channel-dependent parameters and
cutoffs fitted in Ref.~\cite{Duan:2024} and subsequently adopted in
Ref.~\cite{Ikeno:2026zpv}. Their central values are collected in
Table~\ref{tab:interaction_parameters}. Physical charge states belonging to
the same family within this sector receive the same pair of parameters, while
their physical masses are kept in the loop functions. The
$\bar K^0p$ sector instead uses the values
$f_i=1.15f_\pi=106.95$ MeV and $q_{\max,i}=630$ MeV for every channel, following what is done in Refs.~\cite{Oset:1997it,Li:2025yad}.
\begin{table}[t]
\caption{Channel-dependent parameters entering the neutral-sector
Weinberg--Tomozawa interaction and loop functions.}
\label{tab:interaction_parameters}
\begin{ruledtabular}
\begin{tabular}{lcc}
Channel family & $f_i$ (MeV) & $q_{\max,i}$ (MeV)\\
\hline
$\bar KN$      & $42.94$  & $400.95$\\
$\pi\Lambda$   & $40.35$  & $860.65$\\
$\pi\Sigma$    & $179.54$ & $1257.47$\\
$\eta\Lambda$  & $103.29$ & $828.91$\\
$\eta\Sigma$   & $108.40$ & $1043.17$\\
$K\Xi$         & $147.15$ & $1389.43$\\
\end{tabular}
\end{ruledtabular}
\end{table}

For a meson of mass $m_i$ and a baryon of mass $M_i$, the loop function used
in Eq.~(\ref{eq:bethe_salpeter}) is written as
\be
G_i(W)=2M_i I_{\rm cut}(W;m_i,M_i,q_{\max,i}).
\label{eq:loop_definition}
\ee
Defining
\be
s=W^2+i0,\qquad \Delta_m=m^2-M^2,
\qquad
L_m=\sqrt{1+\frac{m^2}{q_{\max}^2}},\qquad
L_M=\sqrt{1+\frac{M^2}{q_{\max}^2}},
\label{eq:loop_definitions}
\ee
and
\be
\sigma(s)=\sqrt{-\left[s-(m+M)^2\right]
                    \left[s-(m-M)^2\right]},
\label{eq:loop_sigma}
\ee
the cutoff scalar loop is
\begin{align}
I_{\rm cut}=\frac{1}{16\pi^2s}\Bigg\{&
\sigma(s)\left[
\tan^{-1}\left(\frac{s+\Delta_m}{\sigma(s)L_m}\right)
+\tan^{-1}\left(\frac{s-\Delta_m}{\sigma(s)L_M}\right)
\right]
\nonumber\\
&-(s+\Delta_m)\ln\left[\frac{q_{\max}(1+L_m)}{m}\right]
-(s-\Delta_m)\ln\left[\frac{q_{\max}(1+L_M)}{M}\right]
\Bigg\}.
\label{eq:loop_analytic}
\end{align}
This analytic cutoff representation follows
Refs.~\cite{Guo:2006,Lyu:2026rsm}.
This unitarized construction produces the low-lying $S=-1$ baryon states
dynamically. In particular, it gives the two-pole structure associated with
the $\Lambda(1405)$ and also poles related with the $\Lambda(1670)$~\cite{Oset:2001cn,Jido:2003cb}.

\subsection{Production followed by rescattering}
\label{subsec:production_fsi}

The scalar $S$-wave production amplitude is organized as one direct term plus
two loop--rescattering corrections,
\be
{\cal M}_s={\cal S}_{\rm LO}F_s,
\qquad
F_s=\widetilde B+
\Delta F_{K^-p}(M_{pK^-})+
\Delta F_{\bar K^0p}(M_{p\bar K^0}).
\label{eq:Fs}
\ee
For the neutral final pair we have
\begin{align}
\Delta F_{K^-p}={}&
\widetilde B G_{K^-p}T_{K^-p,K^-p}
+2\widetilde B G_{\bar K^0n}T_{\bar K^0n,K^-p}
\nonumber\\
&+\widetilde B G_{\pi^+\Sigma^-}T_{\pi^+\Sigma^-,K^-p}
+\widetilde A G_{\pi^-\Sigma^+}T_{\pi^-\Sigma^+,K^-p}
\nonumber\\
&+\frac{\widetilde A+\widetilde B}{2}
G_{\pi^0\Sigma^0}T_{\pi^0\Sigma^0,K^-p}
\nonumber\\
&+\frac{2\widetilde C-\widetilde A-\widetilde B}{2\sqrt3}
G_{\pi^0\Lambda}T_{\pi^0\Lambda,K^-p}
\nonumber\\
&-\frac{\widetilde B}{\sqrt6}
G_{\eta\Sigma^0}T_{\eta\Sigma^0,K^-p}
+\frac{\widetilde B}{\sqrt2}
G_{\eta\Lambda}T_{\eta\Lambda,K^-p},
\label{eq:fsi_kmp}
\end{align}
where every function is evaluated at $M_{pK^-}$. The second term of
$\Delta F_{K^-p}$ in Eq.~(\ref{eq:fsi_kmp}), $\bar K^0n\to K^-p$, is essential
because its source has twice the weight of the direct $K^-p$ channel. For the
charged final pair,
\begin{align}
\Delta F_{\bar K^0p}={}&
\widetilde B G_{\bar K^0p}T_{\bar K^0p,\bar K^0p}
\nonumber\\
&+\frac{\widetilde B-\widetilde A}{\sqrt2}
G_{\pi^+\Sigma^0}T_{\pi^+\Sigma^0,\bar K^0p}
\nonumber\\
&+\frac{\widetilde A-\widetilde B}{\sqrt2}
G_{\pi^0\Sigma^+}T_{\pi^0\Sigma^+,\bar K^0p}
\nonumber\\
&+\frac{\widetilde A+\widetilde B-2\widetilde C}{\sqrt6}
G_{\pi^+\Lambda}T_{\pi^+\Lambda,\bar K^0p}
\nonumber\\
&+\frac{\widetilde B}{\sqrt3}
G_{\eta\Sigma^+}T_{\eta\Sigma^+,\bar K^0p},
\label{eq:fsi_k0p}
\end{align}
with every function evaluated at $M_{p\bar K^0}$.

\begin{figure}[t]
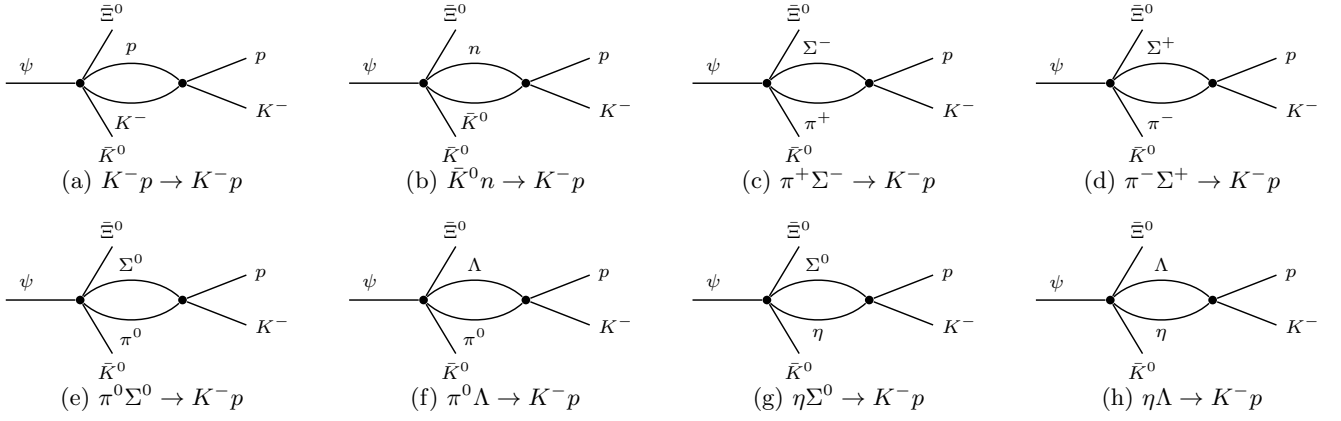

\centering
\begin{minipage}{0.24\linewidth}
\centering
\fsimechanism{K^-}{p}{\bar K^0}{K^-}{p}\\[-2pt]
{\small (a) $K^-p\to K^-p$}
\end{minipage}\hfill
\begin{minipage}{0.24\linewidth}
\centering
\fsimechanism{\bar K^0}{n}{\bar K^0}{K^-}{p}\\[-2pt]
{\small (b) $\bar K^0n\to K^-p$}
\end{minipage}\hfill
\begin{minipage}{0.24\linewidth}
\centering
\fsimechanism{\pi^+}{\Sigma^-}{\bar K^0}{K^-}{p}\\[-2pt]
{\small (c) $\pi^+\Sigma^-\to K^-p$}
\end{minipage}\hfill
\begin{minipage}{0.24\linewidth}
\centering
\fsimechanism{\pi^-}{\Sigma^+}{\bar K^0}{K^-}{p}\\[-2pt]
{\small (d) $\pi^-\Sigma^+\to K^-p$}
\end{minipage}

\medskip
\begin{minipage}{0.24\linewidth}
\centering
\fsimechanism{\pi^0}{\Sigma^0}{\bar K^0}{K^-}{p}\\[-2pt]
{\small (e) $\pi^0\Sigma^0\to K^-p$}
\end{minipage}\hfill
\begin{minipage}{0.24\linewidth}
\centering
\fsimechanism{\pi^0}{\Lambda}{\bar K^0}{K^-}{p}\\[-2pt]
{\small (f) $\pi^0\Lambda\to K^-p$}
\end{minipage}\hfill
\begin{minipage}{0.24\linewidth}
\centering
\fsimechanism{\eta}{\Sigma^0}{\bar K^0}{K^-}{p}\\[-2pt]
{\small (g) $\eta\Sigma^0\to K^-p$}
\end{minipage}\hfill
\begin{minipage}{0.24\linewidth}
\centering
\fsimechanism{\eta}{\Lambda}{\bar K^0}{K^-}{p}\\[-2pt]
{\small (h) $\eta\Lambda\to K^-p$}
\end{minipage}
\caption{Individual meson--baryon rescattering mechanisms that contribute to
the $K^-p$ final-state interaction of Eq.~(\ref{eq:fsi_kmp}). The lower
external $\bar K^0$ line and the upper $\bar\Xi^0$ line are spectators.}
\label{fig:kmp_channels}
\end{figure}

\begin{figure}[t]
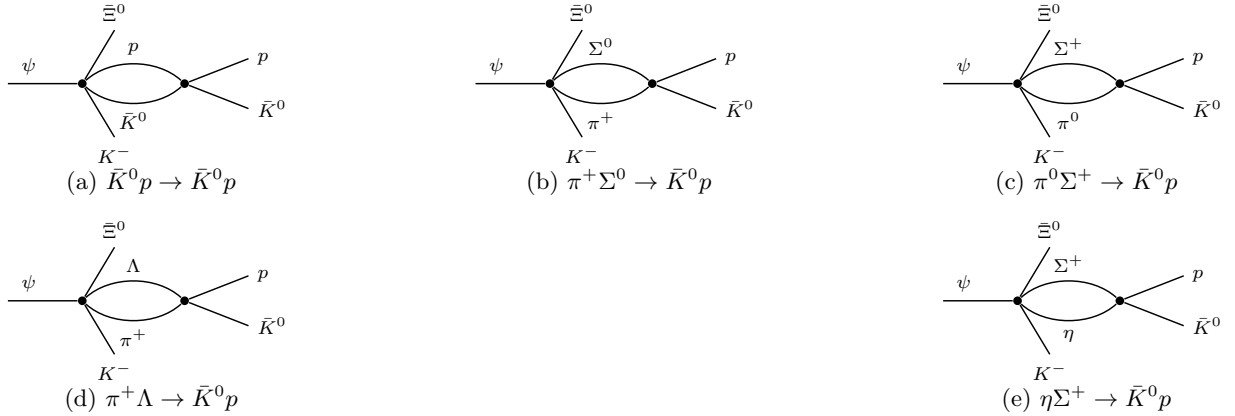

\centering
\begin{minipage}{0.31\linewidth}
\centering
\fsimechanism{\bar K^0}{p}{K^-}{\bar K^0}{p}\\[-2pt]
{\small (a) $\bar K^0p\to\bar K^0p$}
\end{minipage}\hfill
\begin{minipage}{0.31\linewidth}
\centering
\fsimechanism{\pi^+}{\Sigma^0}{K^-}{\bar K^0}{p}\\[-2pt]
{\small (b) $\pi^+\Sigma^0\to\bar K^0p$}
\end{minipage}\hfill
\begin{minipage}{0.31\linewidth}
\centering
\fsimechanism{\pi^0}{\Sigma^+}{K^-}{\bar K^0}{p}\\[-2pt]
{\small (c) $\pi^0\Sigma^+\to\bar K^0p$}
\end{minipage}

\medskip
\begin{minipage}{0.31\linewidth}
\centering
\fsimechanism{\pi^+}{\Lambda}{K^-}{\bar K^0}{p}\\[-2pt]
{\small (d) $\pi^+\Lambda\to\bar K^0p$}
\end{minipage}\hfill
\begin{minipage}{0.31\linewidth}
\centering
\fsimechanism{\eta}{\Sigma^+}{K^-}{\bar K^0}{p}\\[-2pt]
{\small (e) $\eta\Sigma^+\to\bar K^0p$}
\end{minipage}
\caption{Individual meson--baryon rescattering mechanisms that contribute to
the $\bar K^0p$ final-state interaction of Eq.~(\ref{eq:fsi_k0p}). In these
diagrams the $K^-$ and the $\bar\Xi^0$ are spectators.}
\label{fig:k0p_channels}
\end{figure}

Figures~\ref{fig:kmp_channels} and \ref{fig:k0p_channels} display separately
every nonvanishing loop term in Eqs.~(\ref{eq:fsi_kmp}) and
(\ref{eq:fsi_k0p}). Thus, each panel represents the product
$h_iG_iT_{if}$ for the indicated intermediate meson--baryon channel. The
$\bar\Xi^0$ and the second antikaon do not participate in the rescattering
and remain spectators.

\subsection{Separate $\Lambda(1520)$ contribution}
\label{subsec:lambda1520}

The $\Lambda(1520)$ has $J^P=3/2^-$~\cite{ParticleDataGroup:2026mpi} and is not a pole
of the present $S$-wave pseudoscalar--ground-state-baryon matrix. A
possible $\Lambda(1520)\to pK^-$ contribution is indicated by the
BESIII analysis~\cite{BESIII:2026qzk}; we therefore include it as a
separate phenomenological amplitude. Since the $\bar KN$ pair has
negative intrinsic parity, its decay proceeds in the lowest allowed
partial wave $L=2$.

Defining the kaon momentum in the $K^-p$ rest frame as
\be
\widetilde p_K(M)=
\frac{\lambda^{1/2}(M^2,m_K^2,M_p^2)}{2M},
\qquad M=M_{pK^-},
\label{eq:pK_tilde}
\ee
we write the scalar Breit--Wigner part as
\be
F_{1520}(M)=
\frac{\widetilde D\,\widetilde p_K^2(M)/M_p}
{M-M_{\Lambda(1520)}+i\Gamma_{\Lambda(1520)}/2}.
\label{eq:lambda1520}
\ee
Here $\widetilde D$ is a phenomenological production strength and the factor
$\widetilde p_K^2$ gives the required $D$-wave behavior. Up to a normalization
absorbed in $\widetilde D$, the corresponding spin--momentum operator can be
represented as
\be
{\cal S}_{1520}\propto
\epsilon_i\chi_p^\dagger
\left(\widetilde p_{K,i}\,
\bm\sigma\cdot\widetilde{\bm p}_K
-\frac{1}{3}\widetilde p_K^2\sigma_i\right)
\chi_{\bar\Xi}.
\label{eq:lambda1520_spin}
\ee
Ignoring the normalization in Eq.~(\ref{eq:lambda1520_spin}), we can write
\be
\overline{|{\cal M}_{\rm tot}|^2}=
2\left(|F_s|^2+|F_{1520}|^2\right),
\label{eq:total_intensity}
\ee
with $\widetilde D$ to be fitted to the data.

\subsection{Four-body kinematics and phase space}
\label{subsec:phase_space}

We work in the rest frame of the parent,
\be
P_\psi^\mu=(M_\psi,\bm 0),\qquad
p_a^\mu=(E_a,\bm p_a),\qquad
E_a=\sqrt{m_a^2+|\bm p_a|^2},
\quad a\in\{p,K^-,\bar K^0,\bar\Xi^0\}.
\label{eq:rest_frame}
\ee
For an unpolarized decay, the Lorentz-invariant phase-space measure gives
\beqa
d\Gamma&=&\frac{1}{2M_\psi}
(2m_p)(2m_{\bar\Xi^0})\,
\overline{|{\cal M}_{\rm tot}|^2}\,d\Phi_4,
\nonumber\\
d\Phi_4&=&(2\pi)^4\delta^{(4)}\left(P_\psi-\sum_a p_a\right)
\prod_{a=p,K^-,\bar K^0,\bar\Xi^0}
\frac{d^3\bm p_a}{(2\pi)^3 2E_a}.
\label{eq:phase_space_start}
\eeqa
The factor $(2m_p)(2m_{\bar\Xi^0})$ compensates for the normalization
$\bar u u=1$ used for the external baryon spinors.
There are no identical particles in the strong final state, so no symmetry
factor occurs.

We introduce
\be
\bm P_K=\bm p_{K^-}+\bm p_{\bar K^0},\qquad
\bm q=\bm p_{K^-}-\bm p_{\bar K^0}.
\label{eq:sum_difference}
\ee
The inverse transformation and its three-dimensional Jacobian are
\be
\bm p_{K^-}=\frac{\bm P_K+\bm q}{2},\qquad
\bm p_{\bar K^0}=\frac{\bm P_K-\bm q}{2},
\qquad
d^3\bm p_{K^-}d^3\bm p_{\bar K^0}
=\frac18d^3\bm P_Kd^3\bm q.
\label{eq:kaon_inverse_jacobian}
\ee
The factor $1/8$, rather than the one-dimensional value $1/2$, is required
because the transformation is applied independently to the three Cartesian
components.

Let $h=|\bm p_p|$ and choose $\bm p_p=h\hat{\bm z}$. The vector $\bm q$ is
specified in a local system whose $z$ axis is parallel to $\bm P_K$. Writing
$P_K=|\bm P_K|$, $q=|\bm q|$ and $z_q=\cos\widetilde\theta_q$, the two kaon
energies become
\beqa
E_{K^-}&=&\sqrt{m_{K^-}^2+\frac{P_K^2+q^2}{4}
+\frac{P_Kqz_q}{2}},
\nonumber\\
E_{\bar K^0}&=&\sqrt{m_{\bar K^0}^2+\frac{P_K^2+q^2}{4}
-\frac{P_Kqz_q}{2}}.
\label{eq:kaon_energies}
\eeqa
Momentum conservation eliminates the remaining antibaryon momentum, returning
\be
\bm p_{\bar\Xi^0}=-(\bm p_p+\bm P_K).
\label{eq:xi_momentum}
\ee
We define the energy available to the antibaryon as
\be
\Delta=M_\psi-E_p-E_{K^-}-E_{\bar K^0}=E_{\bar\Xi^0}.
\label{eq:Delta}
\ee
If $\theta$ is the angle between $\bm p_p$ and $\bm P_K$, energy conservation
fixes it according to
\be
{\cal A}\equiv\cos\theta=
\frac{\Delta^2-m_{\bar\Xi^0}^2-h^2-P_K^2}{2hP_K}.
\label{eq:A_angle}
\ee
An integration point is physical only when $\Delta>0$ and
$-1\leq{\cal A}\leq1$. The derivative of the energy constraint gives
\be
\int_{-1}^{1}\frac{d\cos\theta}{2E_{\bar\Xi^0}}\,
\delta(M_\psi-E_p-E_{K^-}-E_{\bar K^0}-E_{\bar\Xi^0})
=\frac{\Theta(\Delta)\Theta(1-{\cal A}^2)}{2hP_K}.
\label{eq:angular_delta}
\ee

Considering the rotational invariance of the system and choosing $\bm p_p$ in
the $z$ direction as we have done, the decay width is reduced to five variables,
\begin{align}
\Gamma=\frac{(2m_p)(2m_{\bar\Xi^0})}{2048\pi^6M_\psi}
\int&dE_p\int P_KdP_K\int q^2dq\,dz_q\,d\widetilde\varphi_q
\nonumber\\
&\times
\frac{\overline{|{\cal M}_{\rm tot}|^2}}
{(2E_{K^-})(2E_{\bar K^0})}
\Theta(\Delta)\Theta(1-{\cal A}^2).
\label{eq:phase_space_final}
\end{align}
$z_q$ and $\widetilde\varphi_q$ are the cosine and azimuthal angle of $\bm q$
referred to $\bm P_K$. A rotation is then needed to bring this vector to the
reference frame where $\bm p_p$ is in the $z$ direction. Further details are
given in Ref.~\cite{Song:2022kac}. The integration is done with Monte Carlo. For each accepted random point, the on-shell four-momenta are reconstructed as
\beqa
p_p^\mu&=&(E_p,0,0,h),
\nonumber\\
p_{K^-}^\mu&=&\left(E_{K^-},\frac{\bm P_K+\bm q}{2}\right),
\nonumber\\
p_{\bar K^0}^\mu&=&\left(E_{\bar K^0},\frac{\bm P_K-\bm q}{2}\right),
\nonumber\\
p_{\bar\Xi^0}^\mu&=&\left(\Delta,-\bm p_p-\bm P_K\right).
\label{eq:four_momenta}
\eeqa
All six two-body invariant masses then follow from
\be
M_{ab}^2=(p_a+p_b)^2
=m_a^2+m_b^2+2(E_aE_b-\bm p_a\cdot\bm p_b).
\label{eq:pair_mass}
\ee
In particular, $M_{pK^-}$ and $M_{p\bar K^0}$ are used separately as the
arguments of Eqs.~(\ref{eq:fsi_kmp}) and (\ref{eq:fsi_k0p}).

As the integral of Eq.~(\ref{eq:phase_space_final}) is performed by generating
random values of the variables, the different invariant masses for each
generated point are evaluated, and the strength of the integrand is
accumulated in boxes of the given invariant mass for each pair of particles.
Further details can be seen in Refs.~\cite{Song:2022kac,Yang:2026bhm}.

\section{Numerical results}
\label{sec:results}
\label{sec3}

\begin{figure}[H]
\centering
\includegraphics[width=0.92\textwidth]{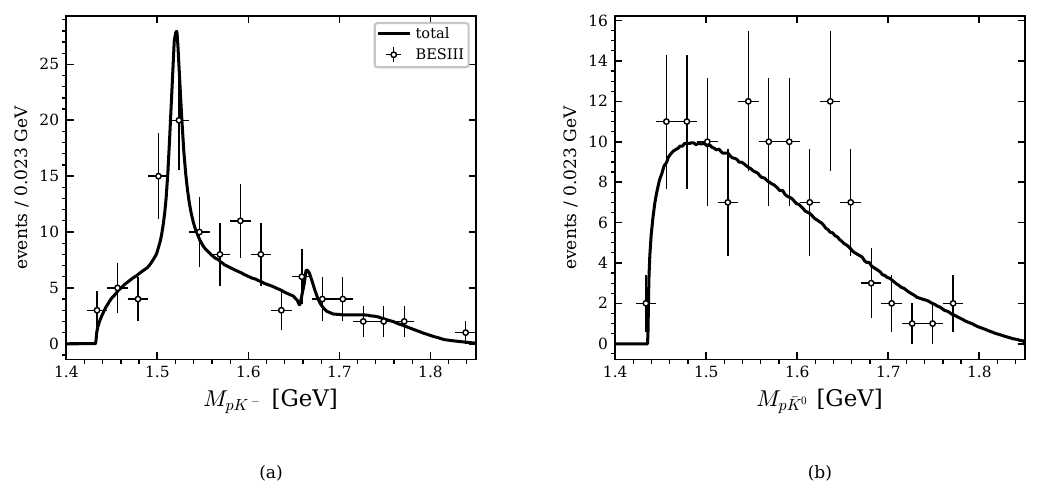}
\caption{The $pK^-$ and $p\bar K^0$ invariant-mass distributions in the
$\psi(3686)\to pK^-\bar K^0\bar\Xi^0$ decay. The solid line is the result of
the simultaneous fit. Experimental data are taken from
Ref.~\cite{BESIII:2026qzk}.}
\label{fig:results_primary}
\end{figure}

\begin{figure}[H]
\centering
\includegraphics[width=0.86\textwidth]{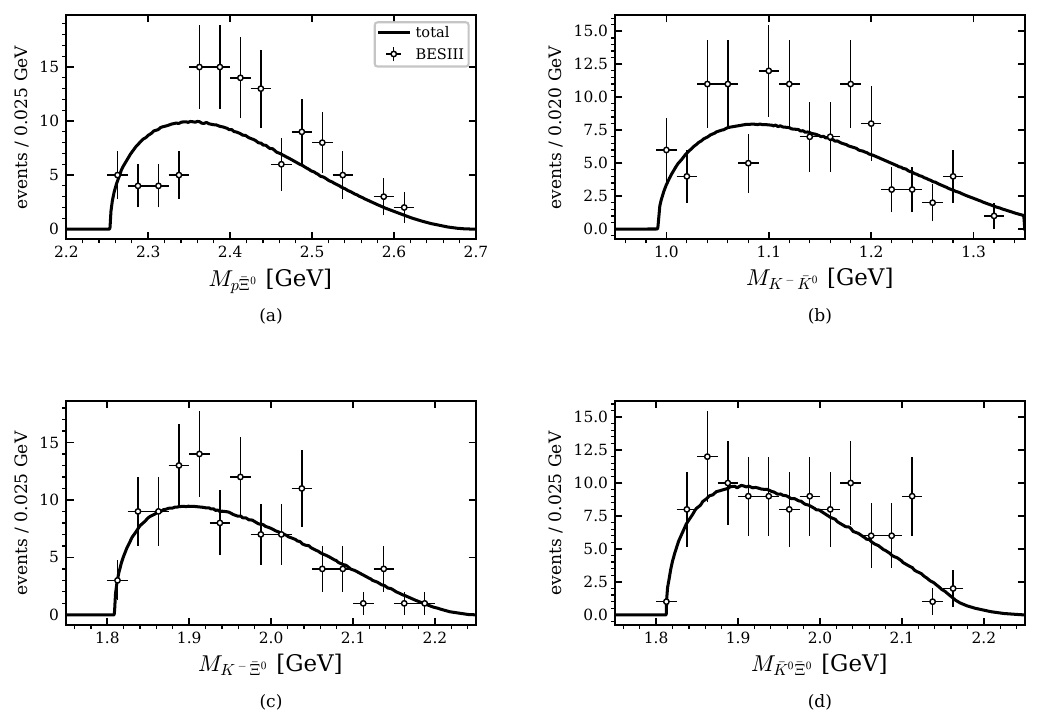}
\caption{The remaining four two-body invariant-mass distributions. The solid
line is the result of the same simultaneous fit as in
Fig.~\ref{fig:results_primary}. Experimental data are taken from
Ref.~\cite{BESIII:2026qzk}.}
\label{fig:results_reflections}
\end{figure}

We show first the two distributions that contain an interacting
proton--antikaon pair in Fig.~\ref{fig:results_primary}. The other four
two-body invariant-mass distributions are shown in
Fig.~\ref{fig:results_reflections}. A single set of parameters is used for all
six spectra. We have the three production strengths $\widetilde A$,
$\widetilde B$, and $\widetilde C$, together with the strength
$\widetilde D$ of the separate $\Lambda(1520)$ contribution. In the
normalization convention of the numerical calculation, the fitted values are
\be
(\widetilde A,\widetilde B,\widetilde C,\widetilde D)
=(4.77\pm2.90,8.76\pm2.15,10.6\pm11.3,2.36\pm0.36)\times10^{-4},
\label{eq:fitted_strengths}
\ee
which correspond to $\widetilde A/\widetilde B=0.544$,
$\widetilde C/\widetilde B=1.21$, and
$\widetilde D/\widetilde B=0.269$.
The fit yields $\chi^2/\mathrm{d.o.f.}=86.28/90\simeq0.959$.
The ratio $\widetilde C/\widetilde B$, however, is only weakly constrained as it is verified by its larger 1$\sigma$ error.
In the prescription used here, $\widetilde C$
enters the calculated spectra only through the $\pi\Lambda$ source
coefficients proportional to
$2\widetilde C-\widetilde A-\widetilde B$, whose numerical contribution is
small. A profile scan, with the remaining parameters refitted at each point,
gives $\Delta\chi^2<1$ over approximately
$0.3\lesssim\widetilde C/\widetilde B\lesssim1.7$; in particular, the
solutions near $0.86$ and $1.21$ differ by only
$\Delta\chi^2\simeq0.1$. Therefore, the quoted central value should not be
interpreted as a precise determination of $\widetilde C/\widetilde B$.

We see in Fig.~\ref{fig:results_primary}(a) a narrow peak around $1520$ MeV,
which corresponds to the $\Lambda(1520)$ in the $pK^-$ system.
There is no analogous narrow structure in the $p\bar K^0$ distribution of
Fig.~\ref{fig:results_primary}(b), since this pair is a pure $I=1$ state. It
therefore selects the isovector $\bar K N$ amplitude, whose near-threshold
enhancement is associated with the dynamically generated
$\Sigma(1430)$~\cite{Li:2025yad}. The physical $\bar K^0p$ threshold is
$1435.88$ MeV, so the position of this state lies at, or slightly
below, the lower kinematic boundary. Only its above-threshold tail is
accessible, and the strength is consequently compressed against the low-mass
end of the physical interval rather than forming a symmetric resonance peak.
A second, much smaller enhancement is visible in
Fig.~\ref{fig:results_primary}(a) around the expected $\Lambda(1670)$ region.
It is generated by the coupled-channel $S$-wave amplitude, rather than being
introduced as an explicit resonance term. Its appearance follows from the
interaction parameters determined in Ref.~\cite{Duan:2024} from the
Belle $K^-p$ spectrum near the $\eta\Lambda$ threshold and
$K^-p\to\eta\Lambda$ data, and subsequently used in
Ref.~\cite{Ikeno:2026zpv} to reproduce the corresponding experimental
invariant-mass distributions in the
$\psi(3686)\to\Lambda\bar\Lambda\eta$ decay. The modest bump obtained here is therefore
consistent with the same dynamical origin. The data seem to show a small bump,
based on a single datum, compatible with our result, but the present
statistical situation does not allow us to elaborate further. It is interesting to mention that the $\Lambda(1670)$ appears as a clear peak in the
$\eta\Lambda$ spectrum in $\psi(3686)\to\Lambda\bar\Lambda\eta$ reaction, whereas here it gives only a small signal in the
$\bar Kp$ channel, as a consequence of the small coupling of this resonance
to $\bar KN$.

The distributions of Fig.~\ref{fig:results_reflections} exhibit broad
structures generated by the available phase space and by kinematic
reflections of the two interacting proton--kaon pairs. The reflection of the
$\Lambda(1520)$ is also present in these projections, but is spread over a
large mass interval. Hence, its appearance outside the $M_{pK^-}$ spectrum
does not imply a resonance in the pair displayed on the horizontal axis.

\begin{figure}[H]
\centering
\includegraphics[width=0.92\textwidth]{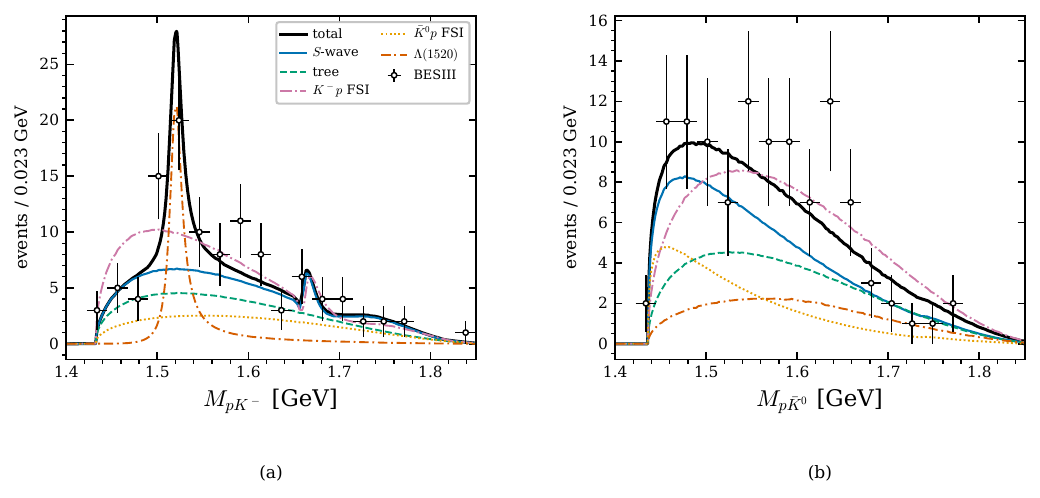}
\caption{Contributions to the $pK^-$ and $p\bar K^0$ invariant-mass
distributions. The solid black line is the total result.}
\label{fig:components_primary}
\end{figure}

\begin{figure}[H]
\centering
\includegraphics[width=0.86\textwidth]{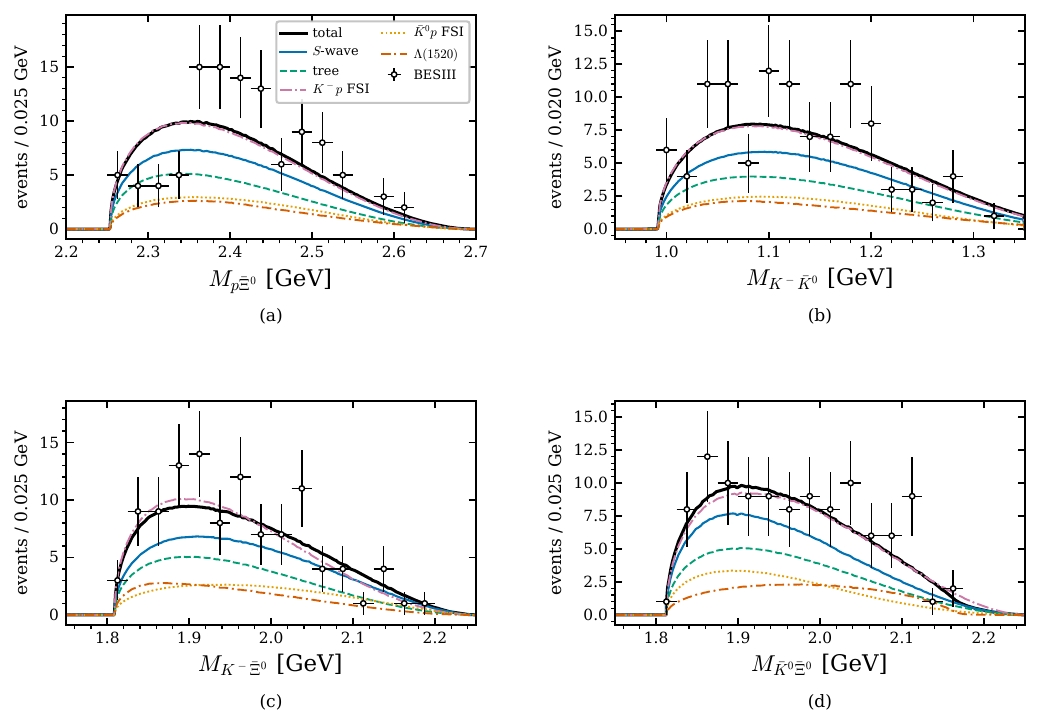}
\caption{Contributions to the four invariant-mass distributions of
Fig.~\ref{fig:results_reflections}. The solid black line is the total result.}
\label{fig:components_reflections}
\end{figure}

In order to understand the relevance of the different terms, we show their
contributions in Figs.~\ref{fig:components_primary} and
\ref{fig:components_reflections}. The curve labeled tree retains only the
direct $\widetilde B$ term in Eq.~(\ref{eq:Fs}), while the curves labeled
$K^-p$ FSI and $\bar K^0p$ FSI retain separately the corresponding loop
term. These are isolated intensities and do not add to the $S$-wave curve,
which contains their coherent sum and all mutual interference terms. The
$\Lambda(1520)$ contribution is added incoherently.

Figure~\ref{fig:components_primary}(a) makes clear that the $\Lambda(1520)$ term
produces the narrow $pK^-$ peak, while the coherent $S$-wave amplitude gives
the broad background underneath it. The rescattering contribution is of the
same order as the tree-level term, and their interference is therefore
essential for the final line shapes. In particular, the small structure near
$1670$ MeV belongs to this coherent $S$-wave contribution. In addition,
Fig.~\ref{fig:components_primary}(b) shows that the $\bar K^0p$ FSI is more
peaked towards threshold than any of the other contributions and is
responsible for the particular shape of this distribution, which differs
substantially from phase space.

The isolated $\Lambda(1520)$ curves in
Fig.~\ref{fig:components_reflections} display the broad replicas generated by
the same events that form the narrow peak in $M_{pK^-}$. They become less
prominent after the coherent $S$-wave contribution is included.

\begin{figure}[H]
\centering
\includegraphics[width=0.92\textwidth]{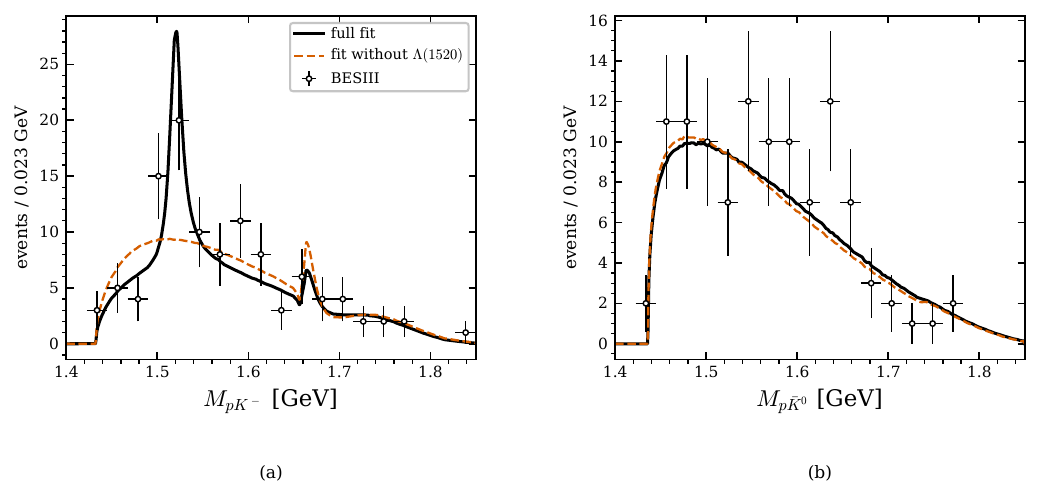}
\caption{Comparison of the full result with the best simultaneous fit obtained
after setting $\widetilde D=0$ for the two proton--antikaon distributions.}
\label{fig:no_lambda_primary}
\end{figure}

\begin{figure}[H]
\centering
\includegraphics[width=0.86\textwidth]{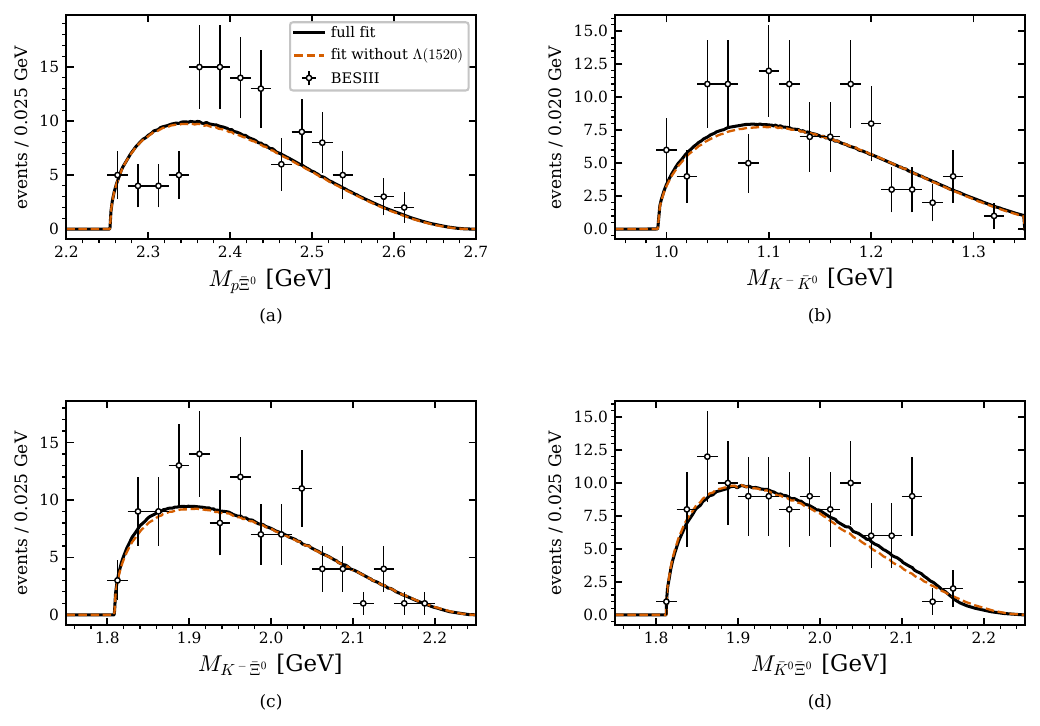}
\caption{The same comparison as in Fig.~\ref{fig:no_lambda_primary} for the
remaining four invariant-mass distributions.}
\label{fig:no_lambda_reflections}
\end{figure}

To further show the role of the $\Lambda(1520)$, we repeat the fit with
$\widetilde D=0$ and readjust the other three strengths. The results for the
interacting pairs are shown in Fig.~\ref{fig:no_lambda_primary}, and those for
the remaining projections in Fig.~\ref{fig:no_lambda_reflections}.

The calculation without the $\Lambda(1520)$ misses the narrow strength around $1520$ MeV in Fig.~\ref{fig:no_lambda_primary}(a), while the other projections change only moderately. This supports the
$\Lambda(1520)$ contribution suggested in the experimental analysis.

The fit uses the exact experimental bin limits and the 94 bins with nonzero quoted statistical uncertainty. With four fitted parameters, the full result has 90 nominal degrees of freedom. The phase-space integration uses $10^8$ Monte Carlo points and the same accepted sample is used for all six projections.

\section{Conclusions}\label{sec4}

We have studied the decay $\psi(3686)\to p K^- \bar K^0 \bar\Xi^0$ within a framework in which SU(3) symmetry is assumed at the production vertex, before final-state interactions take place. We treat the $\psi(3686)$ as a $c\bar c$ state and hence as an SU(3) singlet. We then identify the combinations of a baryon, an antibaryon, and two pseudoscalar mesons that form an SU(3) singlet, retaining all terms that can contribute to the $p K^- \bar K^0 \bar\Xi^0$ final state after final-state interactions.

Using results for meson–baryon interactions from the chiral unitary approach, we show that the relevant final-state interactions are those between a meson and a baryon, with the $\bar\Xi^0$ acting as a spectator. We also justify neglecting the interaction between the two pseudoscalar mesons, which together carry strangeness $-2$. Within this framework, the $\Lambda(1670)$ and $\Sigma(1430)$ resonances—generated in the chiral unitary approach, with the latter recently observed by the Belle collaboration near the $\bar K N$ threshold—produce visible effects. The $K^-p$ invariant-mass distribution also exhibits a clear $\Lambda(1520)$ peak. Since this resonance contributes in the $D$-wave, it does not interfere with the other resonances arising from $S$-wave meson–baryon interactions.

A fit with three parameters and one overall normalization provides a very good description of all six invariant-mass distributions. The $\Lambda(1670)$ contribution to the $K^-p$ spectrum is small, in agreement with the data, and contrasts with the prominent peak seen in the $\eta\Lambda$ invariant-mass distribution in other reactions. This is consistent with the weak coupling of the $\Lambda(1670)$ to the $\bar K N$ channel, a natural outcome of the chiral unitary approach. The $\Sigma(1430)$ appears near the $\bar K N$ threshold and produces an enhancement in the $\bar K^0p$ invariant-mass distribution, leading to a noticeable deviation from phase space.

Overall, our results show that charmonium decays into a baryon–antibaryon pair and two pseudoscalar mesons provide a valuable complement to the more commonly studied decays into a baryon–antibaryon pair and one pseudoscalar meson, offering an additional means of investigating meson–baryon dynamics.

\begin{acknowledgments}
This study was financed, in part, by the S\~ao Paulo Research Foundation
(FAPESP), Brasil. Process Numbers \#2024/19103-9 and \#2026/04916-0.
This work was supported by the National Key R\&D Program of China
(Grant No.~2024YFE0105200), the Natural Science Foundation of Henan
(Grant No.~252300423951), and the Zhengzhou University Young Student
Basic Research Projects for PhD students (Grant No.~ZDBJ202522).
Wen-Tao Lyu also acknowledges support from the China Scholarship Council.
Additional support was provided by Grants PID2023-147458NB-C21 and
CEX2023-001292-S, funded by MICIU/AEI/10.13039/501100011033 and ERDF/EU,
and by the PROMETEO program under Grant CIPROM/2023/59, funded by
Generalitat Valenciana 10.13039/501100003359.
\end{acknowledgments}

\bibliographystyle{apsrev}
\bibliography{references}

\end{document}